\documentclass[twocolumn,preprintnumbers,amsmath,amssymb]{revtex4}
\usepackage{graphicx}
\usepackage{dcolumn}
\usepackage{bm}

\newcommand{\comment}[1]{}
\newcommand{\li}{$^{6}$Li }
\newcommand{\yb}{$^{174}$Yb }

\usepackage{color}

\begin{document}

\title{Quantum Degenerate Mixture of Ytterbium and Lithium Atoms}

\author{Anders H. Hansen}
\author{Alexander Khramov}
\author{William H. Dowd}
\author{Alan O. Jamison}
\author{Vladyslav V. Ivanov}
\author{Subhadeep Gupta}

\affiliation{Department of Physics, University of Washington, Seattle WA 98195}

\date{\today}
\begin{abstract}
We have produced a quantum degenerate mixture of fermionic alkali \li and
bosonic spin-singlet \yb gases. This was achieved
using sympathetic cooling of lithium atoms by evaporatively cooled
ytterbium atoms in a far-off-resonant optical dipole trap. We observe
co-existence of Bose condensed ($T/T_{\rm c} \simeq 0.8$) \yb with $2.3\times10^{4}$
atoms and Fermi degenerate ($T/T_{\rm F} \simeq 0.3$) \li with $1.2\times10^{4}$
atoms. Quasipure Bose-Einstein condensates of up to $3\times10^{4}$ \yb atoms can be
produced in single-species experiments. Our results mark a significant step
toward studies of few and many-body physics with mixtures of alkali and
alkaline-earth-like atoms, and for the production of paramagnetic polar molecules
in the quantum regime. Our methods also establish a convenient scheme for producing
quantum degenerate ytterbium atoms in a 1064nm optical dipole trap.
\end{abstract}
\maketitle

Quantum degenerate elemental mixtures can be used to study a variety
of few- and many-body phenomena, and form the starting point
for creating quantum degenerate dipolar molecules. While
bi-alkali quantum mixtures \cite{modu01,hadz02,silb05,aubi06,tagl08,spie09}
have been produced and studied for about a decade, mixtures of alkali
and electron spin-singlet atoms are a more recent development \cite{nemi09,tass10,okan10,baum11,ivan11}.
By exploiting the difference in mass of the
components, the lithium-ytterbium quantum degenerate mixture may be used to investigate a range
of interesting scientific directions including new Efimov states \cite{dinc06,marc08},
impurity probes of the Fermi superfluid \cite{spie09}, and mass
imbalanced Cooper-pairs \cite{iski08,geze09,tren11}. Furthermore,
unlike the bi-alkali case, mixtures of alkali and alkaline-earth-like atoms
can lead to the realization of paramagnetic polar molecules by combining
the atoms through field-induced scattering resonances. Such molecules
hold great promise for quantum simulation and topological quantum
computing applications \cite{mich06}. They may also be good candidates
for sensitive tests of fundamental symmetries, particularly if one
of the constituents is a heavy atom, such as Yb \cite{huds02}.

\begin{figure}
\includegraphics[angle = 0, width = 0.5 \textwidth] {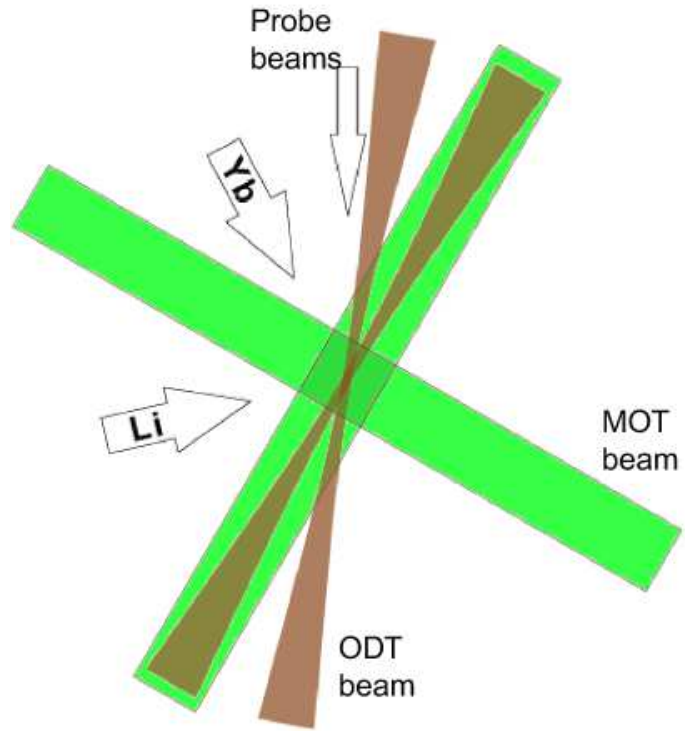}
\vspace{-1cm}
\caption{(color online). Experimental arrangement (top view) for producing
simultaneous quantum degeneracy in lithium and ytterbium. Zeeman slowed
atomic beams of each species propagate along separate axes towards the MOT.
The horizontal ODT beams (brown) are crossed at the MOT region at an angle of
$20^{\circ}$. MOT beams (green) for both species are overlapped and arranged in
retroreflection configuration. Beams for the vertical MOT axis and Zeeman slowing
are omitted in the figure for clarity.}
\label{fig:setup}
\vspace{-0.5cm}
\end{figure}

In this paper, we report on simultaneous quantum degeneracy in a mixture
of alkali and alkaline-earth-like atoms. In earlier work \cite{ivan11}, we
reported on collisional stability and sympathetic cooling in the \li-\yb system, together with a measurement
of the interspecies $s$-wave scattering length magnitude. Here we establish
a convenient method to produce Bose-Einstein condensates (BEC) of \yb\!. This allows
the sympathetic cooling of \li to well below its Fermi temperature and
the achievement of simultaneous quantum degeneracy in the two species.

The cooling of various isotopes of ytterbium to quantum degeneracy
has been pioneered by the group of Y. Takahashi in Kyoto \cite{taka03,fuku07,fuku09}.
In these studies, the optical dipole trap (ODT) was implemented at the wavelength
$532\,$nm. While suitable for confining ytterbium which has a strong
transition at $399\,$nm, this choice of wavelength will not confine
common alkali atoms due to their strong transitions occurring at wavelengths
greater than $532\,$nm. For our ODT, we use $1064\,$nm light arranged
in a straightforward horizontal geometry, and demonstrate efficient evaporative
cooling of \yb to BEC. This establishes a simple setup for studies with quantum degenerate ytterbium gases, particularly in the context of dual-species experiments.

Our experimental setup (see Fig.\ref{fig:setup}) is similar to
what has been described previously \cite{ivan11}. Briefly, we sequentially
load \yb and then \li from respective magneto-optical traps (MOTs)
into the same ODT. We then perform forced evaporative cooling of \yb
by lowering the power in the ODT. This leads to quantum
degeneracy in either single or dual-species experiments. Two improvements
to our earlier setup which are crucial for this work are the use of
higher power in the Yb Zeeman slowing beam resulting in
larger MOT numbers, and the implementation of a tighter ODT geometry \cite{slwrodtnote}
leading to more efficient evaporative cooling.

The ODT is derived from a 1064nm linearly polarized fiber laser,
operated at a power of $45\,$W. In order to control
the trap depth, the output of the laser is sent through an acousto-optic
modulator. The first order output is split into two equal parts
with orthogonal linear polarizations which then propagate horizontally
toward the atoms. Each beam is focused to a (measured) waist of $26\,\mu$m
and the foci are overlapped at an angle of 20 degrees. The trapping potential
is characterized through measurements of trap frequencies by parametric heating. The relative trap depths and frequencies for the two species are $U_{{\rm Li}}/U_{{\rm Yb}}=2.2$ and $\omega_{{\rm Li}}/\omega_{{\rm Yb}}=8.2$.
To monitor atom number and temperature, we quickly switch off the ODT and perform
resonant absorption imaging of both species.

\begin{figure}
\includegraphics[angle = 0, width = 0.5 \textwidth] {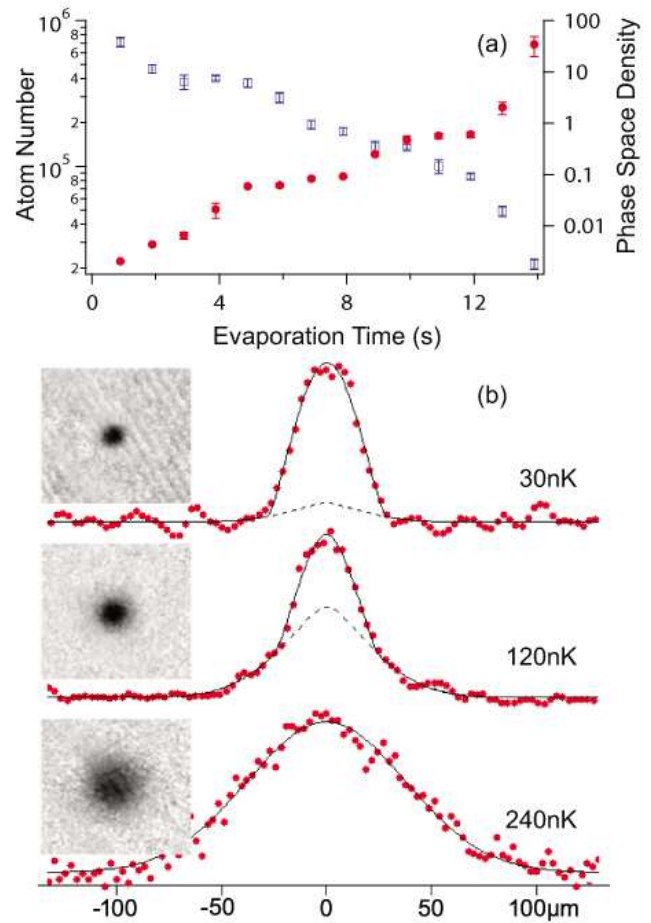}
\vspace{-0.5cm}
\caption{(color online). Evaporative cooling of \yb to Bose-Einstein condensation in the
crossed 1064nm ODT. (a) shows the evolution of \yb number (open squares) and phase space density (filled circles) for a single-species experiment. BEC is achieved after about $12.5\,$s. (b) shows absorption images and the corresponding atomic density profiles (vertical cross-sections of these images) for three different final trap depths, showing the formation of the BEC. The solid line in each plot is a bimodal fit to the distribution with the dashed line showing the thermal component of the fit. The free expansion time after turning off the trap is $8\,$ms for each image. The total atom numbers and temperatures are $(8.0,5.6,2.1)\times 10^4$ and $(240,120,30)\,$nK respectively.}
\label{fig:YbBEC}
\vspace{-0.5cm}
\end{figure}

In single-species experiments with \yb\!, we load $1.5\times10^7$
atoms in a MOT in $40\,$s from a Zeeman-slowed atomic beam. We
use $100\,$mW power in the $399\,$nm (${^1S_0} \rightarrow {^1P_1}$)
slowing beam and a total of $12\,$mW power in the 556nm (${^1S_0} \rightarrow {^3P_1}$)
MOT beams, operated in retro-reflection configuration.
A transient cooling and compression scheme then produces
an atomic cloud at a temperature of $20\,\mu$K containing $\simeq6\times10^{6}$
atoms.

About $1\times10^{6}$ atoms in the ${^1S_0}$ state are
then loaded into the ODT where the background 1/e lifetime is $40\,$s.
The initial ODT power at the atoms is $9\,$W per beam, corresponding to a trap depth of $430\,\mu$K.
The power is then reduced by a factor of 100 over a time scale of $14\,$s, utilizing two stages of
approximately exponential shape. The first stage lasts for $5\,$s with time constant $1.5\,$s.
The second stage lasts for the remainder of the evaporation period and has a time constant of $3.6\,$s.

We observe efficient evaporative cooling with this arrangement (see Fig.\ref{fig:YbBEC}(a)). The critical temperature for Bose-Einstein condensation is achieved after evaporating for $\simeq 12.5\,$s. At this point the atom number is $N_{\rm Yb}=7 \times 10^4$ and the temperature is $T_{\rm Yb}=170\,$nK.
By fitting to the data prior to condensation, we extract an evaporation efficiency
parameter $-{\rm d(ln}(\rho_{\rm Yb}))/{\rm d(ln}(N_{\rm Yb}))=3.4(4)$ where $\rho_{\rm Yb}$ is the phase space density. Nearly pure condensates of up to $3\times10^{4}$ atoms can be prepared by continuing the evaporation process (see Fig.\ref{fig:YbBEC}(b)).

\begin{figure}
\includegraphics[angle = 0, width = 0.5 \textwidth] {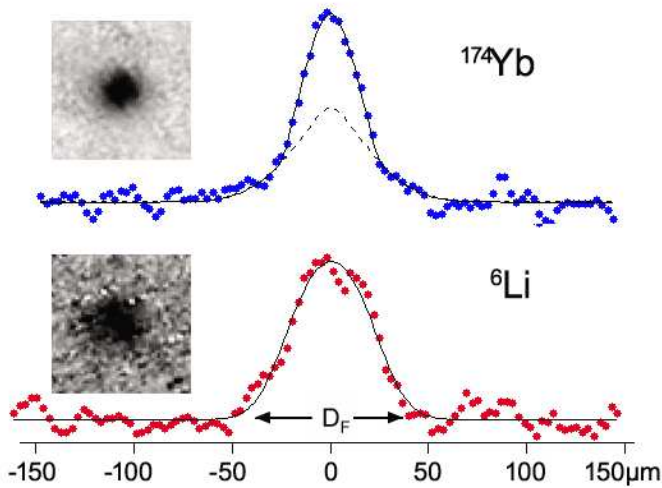}
\vspace{0cm}
\caption{(color online). Quantum degenerate mixture of \yb and \li\!. The absorption images and density profiles correspond to the same experimental iteration with $14\,$s of evaporation. The free-expansion times are $8\,$ms for Yb and $0.7\,$ms for Li. Here $N_{\rm Yb}=2.3 \times 10^4$ and $T_{\rm Yb}=100\,$nK, corresponding to $T_{\rm Yb}/T_{\rm C, Yb}=0.8$, while $N_{\rm Li}=1.2 \times 10^4$ and $T_{\rm Li}=320\,$nK, corresponding to $T_{\rm Li}/T_{\rm F, Li}=0.3$. For \yb\!, the solid line is a bimodal fit with the dashed line showing the thermal component of the fit. For \li\!, the solid line is a Thomas-Fermi fit. The extent of the momentum-space Fermi diameter $D_F$, corresponding to the Fermi energy, is also indicated in the figure.}
\label{fig:double}
\vspace{0cm}
\end{figure}

For two-species experiments, we add to the optically trapped \yb
an equal mixture of the two $F=1/2$ Zeeman states of \li with an
adjustable total number. After $1\,$s of interspecies thermalization
at constant trap depth, we perform sympathetic cooling of \li by
\yb at near-zero magnetic field by using the same evaporation ramp
as described above. Sympathetic cooling works well in this mixture as described in our earlier work \cite{ivan11}
with the \li number remaining nearly constant due to its greater trap depth.
After approximately $14\,$s of evaporation we observe simultaneous quantum degeneracy in the two species (see Fig. \ref{fig:double}). At this point the geometric mean trap frequencies are $\bar{\omega}_{\rm Yb(Li)}=2\pi\times 90(740)\,$Hz, atom numbers are $N_{\rm Yb(Li)}=2.3 (1.2) \times 10^4$, and temperatures are $T_{\rm Yb(Li)}=100(320)\,$nK. Here $N_{\rm Li}$ is the total lithium atom number distributed equally between the two spin states. The difference in temperature between the two species is largely attributable to
the relative center-of-mass displacement at the end of the evaporation ramp arising from gravitational sag. Assuming perfect overlap, the estimated interspecies themalization time at this stage is $\simeq 1\,$s, which is reasonably short. However, the separation of the two clouds due to unequal effects of gravity is $7.5\,\mu$m while the lithium in-trap Fermi diameter is $11.6\,\mu$m in the vertical direction. It is therefore not surprising that sympathetic cooling becomes inefficient towards the end of evaporation.

Our results establish a new quantum system comprised of simultaneously
degenerate one- and two-electron atomic gases. We also demonstrate
a new method for achieving Bose-Einstein condensation of \yb using
a straightforward horizontal optical trapping arrangement with $1064\,$nm laser beams.
Our setup could also be suitable for combining Yb
with other alkalis such as Cs and Rb, since the trap depth and relative sizes would be amenable
for sympathetic cooling by ytterbium. Further improvements to our cooling scheme include independent
control over the powers in the two ODT beams and an additional magnetic field gradient
to improve spatial overlap of the two species.

Extending our method to incorporate alternate ytterbium isotopes
(such as the fermion $^{173}$Yb \cite{fuku07}) appears realistic. This would then realize
Fermi degenerate mixtures with a large mass ratio. Finally, our results
represent a significant milestone toward the production of quantum gases
of paramagnetic polar molecules. Theoretical work on the LiYb molecule has already been
initiated by several groups \cite{zhan10,gopa10,kotonote}. Future experimental work on our system includes photoassociative spectroscopies and searches for Feshbach resonances \cite{zuch10} in this mixture, key steps towards forming the molecule.

This work was supported by the National Science Foundation,
the Alfred P. Sloan Foundation, and NIST. A.K. acknowledges support
from the NSERC. While finishing up this work, we became
aware of similar work \cite{hara11} in which quantum degenerate mixtures
of \li-\yb and \li-$^{173}$Yb were obtained.


\begin{thebibliography}{27}
\bibitem[{slw(2010)}]{slwrodtnote}
\bibinfo{journal}{Compared to our earlier work, the power in the slowing beam
  is about three times larger. For the same ODT laser power, the mean trap
  frequency is also about three times larger.}
\end{thebibliography}
\end{document}